\documentclass[12pt]{article}
\usepackage{amsmath,amssymb,graphicx}
\usepackage{amsthm}
\usepackage{euscript}
\usepackage{eufrak}
\usepackage{slashed}

\newtheorem{lem}{Lemma}
\newtheorem{thm}{Theorem}
\newtheorem{prop}{Proposition}

\newcommand{\pr}{\noindent{\bf Proof}. }
\newcommand{\rem}{\noindent{\bf Remark}. }
\newcommand{\rems}{\noindent{\bf Remarks}. }

\newcommand{\pa}{\partial}
\newcommand{\one}{\cO(1)}
\newcommand{\bpsi}{\bar \psi}

\newcommand{\const}{\textrm{const}}

\newcommand{\hs}{ \hspace{1cm}}

\newcommand{\Vol}{\textrm{Vol}}

\newcommand{\B}{\Big}
\newcommand{\blan}{\Big  \langle} 
\newcommand{\bran}{\Big  \rangle}  
\newcommand{\sgn}{\textrm{sgn}}
\newcommand{\sq}{\square}

\newcommand{\reg}{\textrm{reg}} 
\newcommand{\bJ}{\bar J}

\newcommand{\be}{\begin{equation}}
\newcommand{\ee}{\end{equation}}

\newcommand{\slp}{\slashed{p}} 
 
\newcommand{\slpa}{\slashed{\partial} }

\newcommand{\sZ}{\mathsf{Z}}

\newcommand{\al}{\alpha}

\newcommand{\de}{\delta}
\newcommand{\ga}{\gamma}
\newcommand{\Ga}{\Gamma}

\newcommand{\La}{\Lambda}

\newcommand{\Om}{\Omega}

\newcommand{\ep}{\epsilon}

\newcommand{\vep}{\varepsilon}

\newcommand{\fJ}{\EuFrak{J}}

\newcommand{\cC}{{\cal C}}

\newcommand{\cO}{{\cal O}}

\newcommand{\cR}{{\cal R}}

\newcommand{\cG}{{\cal G}}

\newcommand{\cL}{{\cal L}}

\newcommand{\bbR}{{\mathbb{R}}}
\newcommand{\bbZ}{{\mathbb{Z}}}

\newcommand{\bbT}{{\mathbb{T}}}

\begin{document}

\title{Correlation functions for  the Gross-Neveu model}

\author{J. Dimock\footnote{dimock@buffalo.edu}  \\
Dept. of Mathematics, SUNY at Buffalo \\
Buffalo, NY 14260, USA}

\maketitle

\begin{abstract}    This is  a non-perturbative treatment of  correlation functions for the weakly coupled   massless Gross-Neveu model in a finite volume.   The main  result is that  all correlation functions, treated as distributions,  are uniformly bounded in the ultraviolet cutoff.  
\end{abstract}

\section{Introduction} 

The Gross-Neveu model is a quantum field theory on a two dimensional   spacetime,  here taken to be a torus with  Euclidean metric.   It  describes fermi fields with $n \geq 2$ internal components
and a quartic self interaction  $g \int (\bpsi \psi )^2$.                        In a previous paper \cite{DiYu24} 
 we gave a non-perturbative treatment of the  renormalization group  (RG) flow and used it
to prove bounds on the partition function uniform in the ultraviolet cutoff.   In the present paper  we study correlation functions    $ <\psi(x_1) \cdots \psi(x_n) \bpsi(y_1) \cdots \bpsi(y_n) > $ regarded as distributions.   We find that the established results on the RG flow are sufficient to bound  the correlation functions uniformly in an ultraviolet cutoff (but not in the volume).    

Our results are complementary to earlier results  due to Gawedzki, Kupiainen \cite{GaKu85b}  and  
 Feldman, Magnen, Rivasseau,  Seneor \cite{FMRS86}  as well as subsequent papers referenced in  \cite{DiYu24} .    These papers
 take advantage of the fact that  perturbation theory for the model is convergent.     They expand to all orders in the coupling constant
 and analyze the $n^{th}$ order contribution   with a detailed study of  Feynman graphs.    From this it is possible to extract  information
 about the correlation functions.

We regard both methods as useful.   The present non-perturbative treatment  seems simpler while the complete perturbative treatment
 carries more information in principle.     We should add that the purpose of the present paper is not just  to generate results for the Gross-Neveu model, 
 but also  to develop techniques that    work for models with boson fields  where perturbation theory typically does not converge.

The paper is organized as follows.   In section \ref{preliminaries}  we define the model and recall the renormalization group results.
In section \ref{grassmann} we establish some identities  for Grassmann integration which  play a key role in the analysis.   In section \ref{npoint}
we prove the bounds on the correlation functions.


\section{Preliminaries}       \label{preliminaries} 

We start in a finite volume  by working on the torus $ \bbT_M = \bbR^2/ L^M   \bbZ^2$
 with volume $L^{2M}$.  Here  $L$ is a fixed positive integer and $M$ is 
the long distance cutoff. The fields are elements on an infinite dimensional  Grassmann  algebra generated by  elements 
 $\psi^i _a(x),\bpsi^i _a(x) $ where $x \in \bbT_M$,   the spinor index  is $a  = 1,2$,   and $i=1, \dots,  n$ is the internal index.   
The algebra is constructed in \cite{DiYu24} and  consists of formal expressions of the form 
  \be  
  \begin{split} 
&F(\psi) =  F( \psi, \bpsi)\\
= & \sum_{n,m} \frac{1}{n!m!} \int_{ \bbT^n_M \times \bbT^m_M} F_{nm}(x_1, \dots, x_n,y_1, \dots y_n) \psi(x_1) \cdots \psi(x_n) \bpsi(y_1) \cdots \bpsi(y_n)  dx dy\\
\end{split} 
 \ee
 We usually   have $n=m$. 
 Here the kernels $F_{nm} $ are elements of a Banach space of distributions $\cC'$ which is dual to a Banach space $\cC$ of differentiable functions.  We take
 $\cC = \cC_3( \bbT^{n+m} )$ to be  functions which are three times continuously differentiable  in each variable,   supplied with a sup norm.
 The $F_{nm}$ are anti-symmetric in the $x_i$ and $y_j$ separately.

 A norm  with positive parameters $h,\bar h$ is defined by 
\be \label{norm} 
 \|  F \|_{h, \bar h}  = \sum_{n,m} \frac{h^n\bar h^m }{n!m!}  \|  F_{nm} \|_{\cC'} 
 \ee
 where $\| F_{nm} \|_{\cC'}  = \sup_{ f \in \cC, \|f \| \leq 1} |< F_{nm},f>| $. 
 We usually have $h = \bar h$ in which case the norm is denoted $\|F\|_h$. 
 We define $\cG_h $ to be the  Banach space  of all such objects with $\|F\|_h <\infty$. 
 Multiplication is defined by anti-symmetrizing the kernels, and 
 the  norm satisfies $ \|  F H\|_h \leq  \|  F \|_h  \|  H\|_h $,
 so $\cG_h$ is a Banach algebra.    

 A Gaussian integral with covariance $G(x,y) $ is defined   on monomials by
 \be   \label{cherry2}
  \int \  \psi  (x_1)  \cdots  \psi(  x_n )     \bpsi(y_1) \cdots  \bpsi  (y_m )  \  d \mu_{G}  (\psi)  
  =   \ep_n   \det  \B \{ G (x_i,  y_j)\B  \}   \de_{nm} 
\ee 
with $\ep_n = \pm 1$.  The choice of sign is fixed by specifying  
 \be   \label{cherry1} 
  \int \  \psi  (x_1)  \bpsi(y_1) \cdots  \psi(  x_n ) \bpsi  (y_n  )  \  d \mu_{G}  (\psi )  
  =     \det  \B \{ G (x_i,  y_j)\B  \}  
\ee

  We consider particularly covariances which can be written in the form 
   $G(x,y) = \int G_1(x,z) G_2(z,y) dz$  with
\be \label{shunt} 
h(G)  \equiv \sup_{|\al | \leq 3} \B \{  \|\pa^{\al}G_1\|_2,   \|\pa^{\al}G_2\|_2 \B\} < \infty 
\ee
Then the determinants can be estimated by Gram's inequality,  and one can show that the integral extends to  a  linear functional $ F \to \int F d \mu_G $  on  $\cG_{h(G) } $   
which satisfies 
\be \label{star} 
 | \int F d \mu_G|   \leq  \| F\|_{h(G) } 
 \ee
If $h(G) < h$ then $\cG_h \subset \cG_{h(G) }$  and  the integral is defined on $\cG_h$ with  $ | \int F d \mu_G|   \leq  \| F\|_h$.
\bigskip

The   Euclidean  Dirac operator  is   $\slpa = \sum_{\mu} \ga_{\mu} \pa_{\nu}  = \ga_0 \pa_0 + \ga_1 \pa_1$ 
 where the $\ga_{\mu}$ are  self-adjoint Dirac matrices satisfying  $\{ \ga_{\mu}, \ga_{\nu} \} = \de_{\mu \nu} $.
 We define $G$ to be the inverse Dirac operator with   a momentum cutoff at approximately  $|p| = L^N$.
 It  has a kernel  defined by the Fourier series 
 \be \label{fine} 
G (x-y) =   L^{-2M }  \sum_{ p  \in  \bbT_M^*, p \neq 0}  e^{ip(x-y)}  \frac{-i\slp}{p^2}    e^{-   p^2/L^{2N} }  
 \ee
  where $\bbT_M^* = 2 \pi   L^{-M } \bbZ^2$.

The partition function for the Gross-Neveu model is   a Gaussian integral with  covariance $G$   on the   Grassmann  algebra $\cG_h$  of   the form    
\be \label{initalS} 
\sZ  =
\int  e^{S(\psi) }   d\mu_{G} (\psi)  \hs \hs  S(\psi) =  \int  (    \vep   - z \bpsi \slpa \psi   + g ( \bar \psi \psi )^2  )  dx
\ee
Here    $(\bpsi \psi )(x)  = \sum_{i=1}^{n}  \sum_{a = 1,2} \bpsi^i_{a} (x)\psi^i_a (x) $ with $n \geq 2$.   The parameters in the action $S(\psi)$ are the  coupling constant $g$,  incremental field strength $z$,
and energy density $\vep$.

 Our goal is to get non-perturbative  bounds uniform in the ultraviolet cutoff $N$   as $N \to \infty$.    In an earlier paper this was accomplished for the partition function.
 Now we are interested in accomplishing the same for  correlation functions
 \be  \label{undo} 
 \begin{split}
 & \blan \psi(x_1) \cdots \psi(x_n) \bpsi(y_1) \cdots \bpsi(y_n) \bran \\
 \equiv & \sZ^{-1} \int   \psi(x_1) \cdots \psi(x_n) \bpsi(y_1) \cdots \bpsi(y_n)   e^{S(\psi) }   d\mu_{G} (\psi) \\
\end{split} 
 \ee

It is convenient to scale up to the torus $\bbT_{M+N}$ with a unit momentum cutoff.   
This is accomplished  by introducing $\psi$ on $\bbT_{M+N}$, and  replacing the former   $\psi$ on $\bbT_{M} $ by    $\psi_{L^{-N}  }$   where 
\be \label{ten} 
  \psi_ {L^{- N }}(x)   = L^{ N/2}   \psi( L^{N} x )   \hs      \bpsi_ {L^{- N }} (x)  =   L^{ N/2}   \bpsi( L^{N} x )   
\ee 
We find a new covariance 
\be
G_{0}  (x-y) =   L^{-2(M+N)  }  \sum_{ p  \in  \bbT_{M+N}^*, p \neq 0 }   e^{ip(x-y)} \frac{-i\slp}{p^2}   e^{- p^2 } 
 \ee
and  for the partition function
\be
\sZ =  \int e^{S_0( \psi)  } d \mu_{G_{0 } } ( \psi )  \hs \hs S_0(\psi) =   \int  ( \vep_0  -     z_0 \bpsi  \slpa  \psi  +  g_0   ( \bar \psi \psi )^2   ) dx
\ee
  where the new action has   parameters      $ \vep_0  = L^{-3N} \vep, z_0 = z,  g_0 = g$.
 The  new correlation functions   are
 \be
  L^{Nn} \  \sZ^{-1}      \int   \psi(L^Nx_1) \cdots \psi(L^N x_n) \bpsi(L^N y_1) \cdots \bpsi(L^N y_n)   
  e^{S_0( \psi)  }  d\mu_{G_0} (\psi)  
 \ee

We make a series of RG transformations.  In each step we integrate out the highest momentum modes, approximately those in  $L^{-1} \leq |p| \leq 1$.
Then we scale  the volume down by  $L^{-1} $ and   momentum up by $L$ restoring  the unit momentum cutoff.  
  We  review how this plays out for the partition function, leaving  correlation functions to section \ref{npoint}.
   
  After $k$ steps one has fields $\psi$  on $\bbT_{M+N -k} $   and the expression
\be  \label{start} 
\sZ = \int e^{S_k(\psi) } d \mu_{G_k} ( \psi) 
\ee
The  covariance is  now given by
 \be \label{gk} 
G_k(x - y)  =   {\sum}'_{p  \in \bbT^*_{N+M-k} }    e^{ip(x-y)}  \frac{-i\slp}{p^2}    e^{- p^2 }   
\ee
Here ${\sum}'$ means the sum is weighted by   $L^{ -2(M+N-k)}$ and $p=0$ is excluded. 
  The  effective action $S_k$    is specified below.

Given the representation on $\bbT_{M+N -k} $  we generate the representation  on $\bbT_{M+N -k-1} $ as follows.
 Split off the highest momentum piece of the covariance by 
\be
G_k  =   G_{k+1,L }  + C_k
\ee
where   $G_{k+1,L } (x, y)   = L^{-1} G_{k+1} (L^{-1}x, L^{-1} y )$.   Then 
\be 
\begin{split}  
G_{k+1,L } (x-y)  = &  {\sum}' _{ p \in  \bbT_{M+N-k }^*}  e^{ip(x-y)}  \frac{-i\slp}{p^2}   e^{- L^2p^2  }      \\ 
C_k(x-y)  = &  {\sum}' _{ p  \in  \bbT_{M+N-k }^*}  e^{ip(x-y)}  \frac{-i\slp}{p^2}  \B(  e^{- p^2 } -  e^{- L^2p^2  } \B)     \\
\end{split} 
\ee
The integral splits  as 
\be      \label{original} 
 \sZ =   \int        \B[  \int e^{   S_{k} ( \psi + \eta)} d\mu_{C_k} (\eta) \B]   d \mu_{G_{k+1},L} ( \psi ) 
 \ee
 The expression in brackets is the fluctuation integral,  and a basic problem is to construct  a new action $\tilde S_{k+1}$
 such that
 \be \label{ant} 
 e^{   \tilde S_{k+1}(\psi)    }  =   \int e^{  S_{k} ( \psi + \eta)} d\mu_{C_k} (\eta) 
 \ee
 We scale this by taking  $\psi$ on  $\bbT_{M+N - k -1}$, and replacing   $\psi(x)$ on   $\bbT_{M+N -k }$       by  $\psi_{L } (x) = L^{-\frac12} \psi(L^{-1} x)$. 
 Then $G_{ k+1,L } $ scales to $G_{k+1} $  and  $\tilde S_{k+1} $ scales to 
$S_{k+1}(\psi)   = \tilde S_{k+1} ( \psi _L  ) $, and we have 
\be \label{stop} 
\sZ  =    \int   e^{  S_{k+1} (\psi) }  d \mu_{  G_{k+1}  } ( \psi ) 
\ee

The   effective action after $k$ steps   has the form  
\be \label{bongo} 
\begin{split}
S_k(\psi)   = &  \int  \B(   \vep_k   - z_k  \bpsi \slpa \psi   +  g_k  (\bpsi \psi )^2  \B) +  Q_k^{\reg}(\psi)    +  E_k (\psi)   \\
& +   p_k \int ( \bpsi \ga_5 \psi )^2 +  v_k \int  \sum_{\mu}    (\bpsi \ga_{\mu} \psi )^2  + {Q_k^{\reg}}' (\psi)  \\
\end{split}
\ee
This is established in  \cite{DiYu24}, see particularly section 3.1.  
The various terrms are defined as follows. 
The main term is  $ g_k \int  (\bpsi \psi )^2  \equiv V_k(\psi)$.
It  is again the quartic interaction, now with a new coupling constant $g_k$.   Also new are $\vep_k, z_k = \cO(g_k)$. 

The  $Q_k$ are  $\cO(g_k^2) $ terms  accumulated after $k$ steps.   They have  the form 
\be
 Q_k (\psi)  =\frac12    \int   V^2_k (\psi + \eta)  d \mu_{w_k}(\eta)   -  \frac12 \B(   \int  V _k (\psi + \eta)  d \mu_{w_k}(\eta)  \B)^2 
 \ee
Here  $w_k$ is defined inductively by   $w_{k+1} = (w_k + C_k)_{L^{-1} }$  
and is given explicitly by 
 \be  \label{wk}
w_k(x-y) =  \sum_{j=0}^{k-1} C_{j,L^{-(k-j} }  (x-y) =  {\sum}'_p e^{ip(x-y)} \frac{ -i \slp }{p^2} \B(  e^{-p^2/L^{2k} } - e^{-p^2} \B) 
\ee
The function $w_k(x-y)$ develops short distance singularities $\cO(|x-y|) ^{-1} $ as $k$ increases, and this infects $Q_k$ rendering it singular.
 The   $ Q_k^{\reg}$   is   $Q_k$  with relevant pieces removed     leaving a non singular piece.  The relevant  pieces
 are corrections to the coupling constant, field strength,  and energy density.

 The  $E_k$ is an $\cO(g_k^3) $  element of the Grassmann algebra  of  the form 
\be
E_k(\psi) = \sum_X E_k(X, \psi) 
\ee
where the sum is over  paved sets  $X \subset  \bbT_{M+N -k }$.  A paved set is  a  union of unit blocks (squares),  not necessarily connected. 
The    $E_k(X, \psi)$  only depend on $\psi$ in $X$,  so this expansion localizes the dependence on $\psi$. 
We   employ a norm on   functions $X \to E(X)$ from paved sets to 
 Grassmann elements of the form
  \be
   \| E \|_{h, \Ga} = \sup_{\sq}  \sum_{X \supset \sq} \| E(X) \|_h \Ga(X)
    \ee
  where  $\sq$ is  a unit block.   For translation invariant functions like  $E_k(X)$  the sum  is independent of $\sq$ and the supremum can be omitted.  
  The  weight function
 $\Ga(X) $  grows in the size of the paved set $X$,   so    $ \| E \|_{h, \Ga} < \infty$ enforces that   $E_k(X) $ decays in the size of $X$.  We 
 give  the exact definition  of  $\Ga(X)$  in  Appendix \ref{A}.  We also use  $\Ga_n(X) = e^{n|X|} \Ga(X)$ where $|X|$ is the number of unit blocks in $X$. 

The last three terms in (\ref{bongo})  are rogue   pseudoscalar and vector terms which   do not appear in the original action.
Here $p_k, v_k = \cO(g_k^2) $ and   $ {Q_k^{\reg}}'  = \cO(g_k^3) $  is generated from various quartic terms as above. 
These terms  may actually be zero,  but cannot be ruled out at present.   (Gawedzki and Kupiainen \cite{GaKu85b}  have an argument that such terms do
not occur in a fully developed perturbation theory.    But this does not settle the question in the present context.)

  With this form for $S_k$ an  RG transformation  generates  an action $S_{k+1}$  of the same form but with new parameters.   More precisely we have: 
  
  \begin{thm} \cite{DiYu24}  \label{easter1} 
  Let    $L, h, C_Z, C_E$  be sufficiently large and chosen in that order, and let $g_{\max} $ be sufficiently small depending on these constants.
    Suppose $S_k $ has the form
  (\ref{bongo}) with  
  \be \label{inky} 
   0 < g_k< g_{ \max}  \hs  \|E_k \|_{h, \Ga_4} \leq C_E g_k^3 \hs |z_k| \leq C_Z g_k \hs |p_k|,|v_k| \leq C_Eg_k^2
  \ee
  Then $S_{k+1} $ has the same form with new parameters
  \be  \label{stunning} 
\begin{split}
E_{k+1} = & \cL( \cR E_k + E_k^*) \\
g_{k+1}   =& g_k   +   \beta_k  g^2_k - 2\beta' _kg_kp_k - 4\beta'_k g_kv_k  +  g^*_k \\
z_{k+1}  =&  z_k +  \theta_k  g_k^2- \theta^p_k g_kp_k - \theta^v_k g_kv_k+  z^* _k  \\
p_{k+1}   = &p_k -    2 \beta_k' g_kv_k+ p_k^* \\
v_{k+1}   = & v_k-     \beta'_k g_kp_k+  v_k^* \\
\vep_{k+1} = & L^2 (\vep_k + \vep_k^*) \\
\end{split}
\ee
\end{thm} 

The coefficient $\beta_k>0$  is bounded above and below which ensures ultraviolet asymptotic freedom.  The  other coefficients $ \beta'_k,  \theta_k, \theta^p_k, \theta^v_k$ are bounded. 
The starred quantities  here are all $\cO(g_k^3)$ and have specific estimates.  $\cR E_k$ is $E_k$ with relevant parts removed.  The relevant parts  show up
in the starred quantities.   $\cL$ is a reblocking  and scaling  operation on functions  $E(X, \psi)$. 

The new parameters  $g_{k+1},z_{k+1} , \dots$  do not necessarily satisfy the bounds (\ref{inky}) for $k+1$, but we do have:

 \begin{thm} \cite{DiYu24}  \label{easter2}  Assume  the hypotheses of theorem \ref{easter1}  and suppose  $g_f<  \frac12g_{\max} $.  Then the flow equations (\ref{stunning}) 
   have a unique solution   $\{ g_k, z_k, E_k,   p_k, v_k, \vep_k \}_{0  \leq k \leq N} $ satisfying the bounds  (\ref{inky}) for all $k$ 
with  boundary conditions
\be \label{BC} 
  g_N = g_f,      \  \vep_N =0   \ \ \textrm{ and } \ \   E_0 =0, \     z_0=0, \   p_0=0, \  v_0 =0 
 \ee
\end{thm} 
This is renormalization,  and   we  assume hereafter that we are talking about this solution. 
After $N$ steps  we are back on the original torus $\bbT_M$ and 
the partition function is given by 
\be
\sZ  =    \int   e^{  S_{N} (\psi) }  d \mu_{ G_N   } ( \psi ) 
\ee
The UV singularities have disappeared from  this expression and in \cite{DiYu24}  it is proved that this is bounded uniformly in 
$N$.

 \section{Properties of the Grassmann algebra}  \label{grassmann} 
 
 We digress to   develop some further properties of  our Grassmann  algebra.   Most of these are well-known  in the finite dimensional case,
 but need fresh  proofs in our infinite dimensional setting.    A good  general reference is  \cite{FKT00}, which however does not exactly cover the case at hand.

 \subsection{derivatives}  \label{derivatives}   Again  an element of the Grassmann algebra $\cG_h$ is 
written as
 \be \label{repeat}
  \begin{split} 
 F= & \sum_{n,m} \frac{1}{n!m!} \int_{\bbT^{n+m} }   F_{nm}(x_1, \dots, x_n,y_1, \dots y_m) \psi(x_1) \cdots \psi(x_n) \bpsi(y_1) \cdots \bpsi(y_m)  dx dy \\
\end{split} 
 \ee
 We generalize a bit.  For this section  $ \bbT = \bbT_M = \bbR^d/ L^M   \bbZ^d$    is  any $d$-dimensional  torus, and  the kernels $F_{nm}$   are elements of $\cC'= \cC_k'(\bbT^{n+m} )  $ now the dual space of 
 the Banach space  $\cC = \cC_k(\bbT^{n+m} )$, not just $k=3$.  Here $ \cC_k(\bbT^n )$
 is defined to be all functions  on $\bbT^{n}$ which  are $k$-times continuously differentiable functions in each of the $n$ variables.  The norm is 
 \be
   \| f  \|_{\cC}  = \sup_{ | \al| \leq k } \|\pa^{\al} f\|_{\infty} 
 \ee   
 Here  $\pa^{\al} = \pa^{\al_1} _{x_1}\cdots \pa^{\al_n}_{x_n} $  with $\pa^{\al_i}_{x_i}   = \prod_{\mu=1}^d \pa^{ \al_{i,\mu} }_{ x_{i,\mu} } $ and    $| \al | = \sup_{i} \sum_{\mu}   |\al_{i, \mu} |$.   Compare  the more common  $\cC^k(\bbT^n )$  which would have  $|\al | = \sum_{i, \mu} |\al_{i, \mu} |$.   For $n=1$ we have $\cC_k(\bbT)= \cC^k(\bbT)$.  
 
 The norm $\| F \|_{h, \bar h} $ or $\| F\|_h = \|F\|_{h,h}$ is still defined by  (\ref{norm}).  

 \bigskip

The (formal) derivative  of $F$  with respect to $\psi(x) $ is defined   by $ \pa \psi(x') /\pa  \psi(x) = \de(x-x') $ and    $ \pa \bpsi(x')/\pa  \psi(x) = 0$ and it anticommutes
with fields when distributed across a product.  So the definition is
\be
  \begin{split} 
\frac{ \pa  F } {\pa \psi(x) }  
= & \sum_{n,m} \frac{1}{n!m!}   \sum_{j=1}^n  (-1)^{j+1}  \int  F_{n m}(x_1,  \dots   x_n ,y_1, \dots y_m) |_{x_j =x}  \\
&  \psi(x_1) \cdots \widehat{ \psi(x_j) } \cdots  \psi(x_n) \bpsi(y_1) \cdots \bpsi(y_m)  d \hat x dy \\
\end{split} 
\ee
where $d \hat x = dx_1 \cdots \widehat{ dx_j} \cdots dx_{n+1} $, and the hat indicates omission. 
This is meant to  define  a $\cC'_k(\bbT) $ distribution with values in the algebra.  The precise meaning is that for $f \in \cC_k(\bbT)=\cC^k(\bbT)  $
\be
\begin{split}
&\int f(x)   \frac{ \pa  F}{ \pa \psi(x) }   dx     =\sum_{n,m} \frac{1}{n!m!}  
 \sum_{j=1}^{n}  (-1)^{j+1}  \int   F_{n m}(x_1,  \dots   x_n,y_1, \dots y_m) |_{x_j =x}   \\
&\hs \hs \hs  \psi(x_1) \cdots f(x)  \cdots  \psi(x_n) \bpsi(y_1) \cdots \bpsi(y_m)   d  x dy \\
\end{split} 
\ee
where the $f(x) $  appears  in the $j^{th} $ entry.  The norm of the derivative is  defined as the supremum over $\|f \|_{\cC^k} \leq 1$ of the norm of this expression.  
Derivatives $\pa / \pa \bpsi(x) $ are defined similarly.  
\bigskip

\begin{prop}  \label{first} 
If $F \in \cG_h$  then   $   \pa  F/ \pa   \psi    $  and $    \pa  F/ \pa   \bpsi    $ define  $\cG_{h'} $ valued distributions  for any $h' < h$ and 
\be
\|     \frac{ \pa  F}{ \pa   \psi } \|_{h'}    \leq  \frac{1}{h-h'}     \| F \|_{h  }   \hs   \|     \frac{ \pa  F}{ \pa   \bpsi } \|_{h'}    \leq   \frac{1}{h-h'}     \| F \|_{h }
\ee
\end{prop} 
\bigskip

\pr  First suppose   there are no $\bpsi$ fields, that is $m=0$. 
We have
\be
\|     \frac{ \pa  F}{ \pa   \psi } \|_{h'}  \equiv  \sup_{\| f\|_{\cC^k} \leq 1} \| \int f(x)   \frac{ \pa  F}{ \pa  \psi(x) }    dx     \|_{h'} 
\ee
where
\be
\begin{split} 
&\int f(x)   \frac{ \pa F }{ \pa  \psi(x) }   dx  
=   \sum_{n =1}^{\infty}  \frac{1}{n!}   \sum_{j=1}^{n} (-1)^{j+1}  \int   F_{n}(x_1,  \dots   x_n ) |_{x_j =x}  \psi(x_1) \cdots f(x)  \cdots  \psi(x_n)  d  x  \\
\end{split} 
\ee
In $F_n$ move the $x$ to the left at a cost $(-1)^{j+1} $ and then relabel $ ( x_1,  \dots, x_{j-1}, x_{j+1} \dots x_n)$ as $(y_1, \dots y_{n-1}) $. Then  
\be
\begin{split} 
&\int f(x)   \frac{ \pa F }{ \pa  \psi(x) }   dx  
=   \sum_{n =1}^{\infty}  \frac{1}{(n-1) !}  \int   F_{n}(x, y_1,  \dots   y_{n-1}  ) f(x)   \psi(y_1)   \cdots  \psi(y_{n-1} )  dy  \\
\end{split} 
\ee
This says
\be
\B( \int f(x)   \frac{ \pa  F}{ \pa  \psi(x) }   dx   \B) _{n-1} (y_1, \dots y_{n-1}) =    \int   F_{n}(x, y_1,  \dots   y_{n-1}  ) f(x)  dx 
\ee
With  $\| f\|_{\cC^k} \leq 1$ this implies 
\be
\| \B( \  \frac{ \pa F }{ \pa   \psi}     \B) _{n-1} \|_{\cC'}   \leq      \|  F_n \|_{\cC'} 
\ee
Then 
\be 
\| \frac{ \pa F }{ \pa   \psi}  \|_{h'}  \leq  \sum_{n=1}^{\infty}   \frac{{h'} ^{n-1}}{(n-1) !}   \|  F_n \|_{\cC'}   = \frac{d}{d{h'} }   \| F \|_{h'}
\ee
But  since  $F \in \cG_h$ the function   $h' \to   \| F \|_{h'}$ extends to a complex analytic function for $|h'| < h$.     Then by a Cauchy inequality 
\be
 \frac{d}{d{h'} }   \| F \|_{h'} \leq   \frac{1}{h-h'}     \| F \|_{h} 
\ee
which gives the result in this  case.  

In the general case with both $\psi, \bpsi$ the same argument would give
\be
\|     \frac{ \pa  F}{ \pa   \psi } \|_{h', \bar h'}   \leq \frac{\pa} {\pa h'}  \| F \|_{h', \bar h'  }  
 \leq  \frac{1}{h-h'}     \| F \|_{h, \bar h'  }  \leq     \frac{1}{h-h'}     \| F \|_{h, \bar h } 
 \ee
Then we would specialize to $ \bar h'=h'$ and $ \bar h=h$ to  complete the proof for $\psi$ derivatives.   Derivatives in $\bpsi$ are 
treated similarly.  $\blacksquare$
\bigskip

For the next result we define $F \in \cG_h $ to be even/odd  if it has an even/odd number of fields, i.e. if in the representation (\ref{repeat}) only terms
with $n+m$ even/odd contribute.

\begin{prop} { \ } \label{two} 
\begin{enumerate}
\item
If $F,H \in \cG_h$ and   $F$ is even/odd then
\be
 \frac{ \pa }{ \pa \psi(x) }FH = \B(  \frac{ \pa F }{ \pa \psi(x) } \B) H \pm  F\B(  \frac{ \pa H  }{ \pa \psi(x) } \B) 
\ee
\item If $F \in \cG_h$ is even 
\be
\frac{ \pa F^n  }{ \pa \psi(x) } =n  F^{n-1} \frac{ \pa F  }{ \pa \psi(x) } 
\ee
\item    Suppose $h' < h$ so   $\cG_{h}  \subset \cG_{h'} $.   Let  $ F = \sum_{n=0}^{ \infty} a_n A^n$  be an absolutely   convergent power series  in an even element of  $A \in \cG_h$.  
Then  $\pa F/ \pa \psi(x)  $ in $\cG_{h'} $ is 
\be \label{fluff}
\frac { \pa F }{ \pa \psi(x) } =  \sum_{n=1} ^{\infty} a_n n A^{n-1} \frac { \pa A }{ \pa \psi(x) } 
\ee
\end{enumerate} 
\end{prop} 
\bigskip

\pr  The first follows from the definition of the derivative.  The second follows   by induction on $n$, starting with $n=1$.   

Now consider the third statement.   If $A$ is constant it is trivial.   Otherwise we have $\| A\|_{h'} < \|A\|_h$.   
The series (\ref{fluff}) converges in $\cG_{h'} $ since by Proposition  \ref{first} 
\be 
\|a_n n A^{n-1} \frac{ \pa A }{ \pa \psi } \|_{h'}  \leq \frac{1}{h-h'} | a_n |n \| A\|^{n-1}_{h'} \| A\|_h  =   \frac{1}{h-h'} | a_n | \| A\|^n_{h} 
\left [  n \left( \frac{ \|A\|_{h'}  }{  \|A\|_{h}  } \right)^{(n-1) }   \right] 
\ee
and the bracketed expression is eventually less than one. 
 To establish the identity take the partial sum 
$F_N =  \sum_{n=0}^{N} a_n A^n $ which satisfies 
\be  \label{always} 
\frac { \pa F_N  }{ \pa \psi(x)  }  =  \sum_{n=1} ^{N} a_n n A^{n-1}  \frac { \pa A }{ \pa \psi(x) } 
\ee
As $N \to \infty$ the right side converges to the infinite sum.   The left side  converges to   $\pa F / \pa \psi(x) $
since  again   by Proposition \ref{first}
 \be
\| \frac { \pa F}{ \pa \psi }-\frac { \pa F_N }{ \pa \psi}   \|_{h'} =    
\| \frac { \pa }{ \pa \psi }(F-F_N)  \|_{h'}  \leq    \frac{1}{h-h'}   \|  F- F_N\|_{h} \to 0 
\ee   

 \bigskip
 
\rems   Here are some  examples  of interest,  always assuming $A \in \cG_h$ is even.    If  $e^A$ is defined by the usual power series 
\be   \label{49} 
\frac { \pa e^A }{ \pa \psi(x) }  = e^A\frac { \pa A }{ \pa \psi(x) } 
 \ee
 If  $ \| A\|_h  < 1$  define   
 \be  \label{sunny}   \log (1 +  A ) = \sum_{n=1}^{\infty} \frac{(-1)^{n-1} }{n} A^n
 \ee
and then  
\be \label{sunny2} 
   \frac{ \pa }{ \pa   \psi(x) } \log(1+ A)   
       =   (1+A)^{-1}    \frac{ \pa  A}{ \pa    \psi(x) }          
\ee
We also have
\be \label{sunny3} 
 e^{\log(1+A) } = 1+A
\ee
 To see this define $f(t) =  e^{\log(1+tA) } - (1+tA)$ and check  $f(0) =0,f'(0) =0, f''(t) =0$.

 \subsection{integrals} 
 An  integral on our  Banach algebra $\cG_h$   is a continuous linear functional. 
 Integrals are written   $F \to \int F d\mu$ (although there is no measure). 
 Let $\fJ(x,y)$ be a smooth function on $\bbT \times \bbT$ and define  
 \be < \bpsi, \fJ  \psi> = \int \bpsi(x) \fJ(x,y) \psi (y) dxdy
 \ee
 Then  $ \| < \bpsi, \fJ \psi> \|_h  \leq h^2\| \fJ \|_{\cC'}< \infty $ so this is an element of  $\cG_h$ for any $h$.  The same is true 
 for  $\exp ( \pm < \bpsi, \fJ \psi>  )$ with norm bounded by $ \exp ( \| < \bpsi, \fJ \psi> \|_h) $.  So for any integral on  any $\cG_h$ we can define
 \be
  \int e^{ -< \bpsi , \fJ \psi>   }  d\mu (\psi)  
 \ee
 
 Let  $G$ be a smooth function on $\bbT \times \bbT$ (perhaps taking advantage of an ultraviolet cutoff), and suppose we have  an assignment of $h(G) $ as in (\ref{shunt}).  Then the Gaussian integral $\int F d \mu_G$
 is defined for  $F \in \cG_{h(G)}$ and it is shown in \cite{DiYu24}  that 
 \be
   \int e^{ -< \bpsi , \fJ \psi>   }  d\mu _G(\psi)  = \det( I + \fJ G) 
   \ee 
   On the right $G$ is the operator $(Gf)(x) = \int G(x, y) f(y) dy$. 
The determinant is the Fredholm determinant which exists since $\fJ G $ is trace class on $L^2(\bbT) $.   If $ h(G)<h$ then $\cG_h \subset \cG_{h(G)}  $ and the integral
is also  defined on $\cG_h$ and satisfies the identity.

 Conversely, in the same circumstance,  it is shown in \cite{DiYu24} that if $\int F d \mu$ is an integral on $\cG_h$ satisfying
  \be
   \int e^{ -< \bpsi , \fJ \psi>   }  d\mu (\psi)  = \det( I + \fJ G) 
   \ee 
then   it is  the Gaussian integral  with covariance $G$.  Thus this integral is in a certain sense a   characteristic function of  
the  Gaussian integral. \bigskip

Assume  $F \to \int F d \mu_G$ is a Gaussian integral on $\cG_h$ as above.   Then we have:
 
 \begin{prop}  \label{three} Let  $K(x,y) $ be a smooth function on     $\bbT \times \bbT$  and $K$ the associated  bounded operator on  $L^2(\bbT)$.
 Suppose $\|KG \|  < 1$ so  $(I-KG)^{-1} $ exists.   Then 
  \be
\int  F d\mu_{G,K}(\psi)     \equiv  \frac{  \int  F  e^{<\bpsi, K \psi> }d \mu_G (\psi) }  {  \int   e^{<\bpsi, K \psi>} d \mu_G(\psi) } 
\ee
 is the  Gaussian integral on $\cG_h$ with covariance  $G( I -KG )^{-1} $. 
 \end{prop}
 
 \pr   The denominator is  $\det ( I - KG)$ and it is not zero under the assumption $\|KG \|  < 1$  (Theorem 3.5 in  \cite{Sim79} ).
 The numerator is defined on  $\cG_h$ since   $  e^{<\bpsi, K \psi> } \in \cG_h$.    So it suffices to check 
 that the characteristic function is      $\det ( I + \fJ G( I-K G)^{-1} )$.
We have 
\be 
\begin{split}
 \int e^{ -< \bpsi , \fJ \psi>   }   d\mu_{G,K}(\psi)  
 = &    \frac{  \int    e^{ -<\bpsi,(\fJ- K)  \psi> }d \mu_G (\psi) }  {  \int   e^{<\bpsi, K \psi>} d \mu_G(\psi) } \\
  = &    \frac{  \det ( I  + (\fJ- K) G) }  { \det ( I - KG)  } \\
   = &  \det \B( I + \fJ G( I-K G)^{-1} \B) \\
 \end{split}
 \ee

 \subsection{auxiliary fields}

 Next we    include some auxiliary Grassmann fields   $J(x) = \{ J_a^i(x) \}$ and $ \bJ(x) =\{ \bJ_a^i(x) \}$.   Together        $\psi(x) , \bpsi(x),J(x) , \bJ(x),  $
generate a new Grassman algebra.   The general element has the form
 \be
 \begin{split} 
&  F( \psi ,J ) \\
&=\sum_{n,m,k,\ell }   \frac{1}{ n! m! k! \ell !}  \int_{   \bbT^n \times \bbT^m  \times \bbT^k \times \bbT^{\ell}} 
 F_{ nm k \ell}( x,y,z,w)  \psi(x) \bpsi ( y )    J (z) \bJ(w)   dx dy dz dw \\
 \end{split} 
 \ee 
 but usually we have $n=m$ and $k = \ell$. 
 Here  if  $x= (x_1, \dots x_n)  \in \bbT^n $ then $\psi(x) = \psi(x_1) \cdots \psi(x_n)  $, etc. 
 The kernels $F_{nmk\ell} $ are anti-symmetric  in each of the variables $x,y,z,w$.    A norm is now defined by 
 \be
  \| F \|_{h, h'} =\sum_{k, \ell,n,m }   \frac{h^{n+m} h'^{k + \ell} }{n!m! k! \ell! }  
\| F_{ nm k \ell} \|_{\cC'} 
 \ee 
We define  $\cG_{h,h'} $ as the Banach algebra  of all such objects with  $ \| F \|_{h,h'} < \infty$   

A  Gaussian  integral  over $\psi$  yields an object in the algebra generated by $J, \bJ$ only and  is defined as in (\ref{cherry2})  by 
\be
\begin{split} 
& \int  F( \psi,J ) d \mu_G(\psi)   \\
= &    \sum_{n,m,k,\ell }   \frac{1}{ n! m! k! \ell !}  \int 
 F_{ nm k \ell}( x,y,z,w) \B[ \int  \psi(x) \bpsi ( y ) d \mu_G( \psi)   \B]   J (z) \bJ(w)   dx dy dz dw \\ 
 = &    \sum_{n,m,k,\ell }   \frac{\de_{nm} \ep_n }{ n! m! k! \ell !}  \int 
 F_{ nm k \ell}( x,y,z,w)  \det \{ G(x_i,y_j)  \}   J (z) \bJ(w)   dx dy dz dw \\ 
\end{split}
\ee

For the rest of this section we continue to assume $G$ on $\bbT \times \bbT$ is a smooth function with an assignment
$h(G) $ as in  (\ref{shunt}).

\begin{prop} \label{four} 
If  $F \in \cG_{h(G) ,h'} $  then
$\int  F( \psi,J ) d \mu_G(\psi)   $ exists and 
\be
\|  \int  F( \psi,J ) d \mu_G(\psi) \|_{h'}  \leq  \| F \|_{h(G) ,h'}
\ee
\end{prop} 
\bigskip

\pr   The integral  has the kernel     
\be
F_{k\ell} (z,w) =   \sum_{n,m }   \frac{\de_{nm} \ep_n}{ n! m! }    \int 
 F_{ nm k \ell}( x,y,z,w)  \det \{ G(x_i,y_j)  \}     dx dy
\ee
and the norm
\be \label{pinch}
\|  \int  F( \psi,J ) d \mu_G(\psi) \|_{h'}  =\sum_{k, \ell }  \frac{  h'^{k + \ell} }{  k! \ell! }  
\| F_{ k \ell} \|_{\cC'} 
\ee
By Grams inequality as in  Lemma 7 in \cite{DiYu24}:
 $ 
\| \det \{ G(x_i,y_j)  \} \|_{\cC}  \leq    h(G)^{n+m } 
$.
Hence 
\be
\begin{split} 
\|  F_{k \ell}  \|_{\cC'} &
  = \sup_{ \|f\|_{\cC} \leq 1 }\B| \blan F_{k\ell}, f \bran\B|  \\
&
 \leq  \sup_{ \|f\|_{\cC} \leq 1 }  \sum_{n=m }   \frac{1}{ n! m! } \B| \int 
 F_{ nm k \ell}( x,y,z,w)  \det \{ G(x_i,y_j)  \}  f( z,w)    dx dy \B| \\
 &
  \leq    \sum_{n=m }   \frac{ h(G)^{n+m} }{ n! m! }   \|  F_{ nm k \ell} \|_{\cC'}  \\
 \end{split} 
 \ee
 Put this in (\ref{pinch}) and get the result.  $\blacksquare$
 \bigskip 

  Next note that  $ < \bar J, \psi> $ or  $< \bpsi, J >  $ have kernels $\de (x-y)$ which has $\cC'$ norm $\Vol(\bbT)$.  
Therefore   $ \| < \bar J, \psi> \|_{h,h'}  = h h'\Vol(\bbT)$, and $ < \bar J, \psi> $ or  $< \bpsi, J >  $      are in $\cG_{h,h'}$ for any $h,h'$ as are the exponentials . 
 For any integral on $\cG_{h,h'} $ we can define the generating function  in $\cG_{h'}$  (probably also a characteristic function) by 
 \be
   \int e^{ < \bar J, \psi>  + < \bpsi, J > }  d\mu(\psi) 
 \ee 
 This includes  our Gaussian integrals $\int [ \cdots]d\mu_G$   if $h > h(G) $. 
  
 \begin{prop}  \label{five} 
   \be
  \int e^{ < \bar J, \psi>  + < \bpsi, J > }  d\mu_G(\psi)   =e^{ <\bar J, G J>} 
 \ee
  \end{prop} 
 \bigskip
 
 \pr 
 We have  
 \be
 \begin{split}
 &  e^{ < \bJ, \psi>  + < \bpsi, J > } 
 = \sum_{n, m}  \frac{1}{n!m!}   < \bJ, \psi> ^n   < \bpsi, J >^m  \\
& =   \sum_{n, m}  \frac{1}{n!m!} \int \psi(x_1)\cdots \psi(x_n)  \bpsi(y_1) \cdots \bpsi(y_m)  \bJ(x_1)\cdots \bJ(x_n)  J(y_1) \cdots J (y_m) dx dy\\
 \end{split}
 \ee
 The integral selects $n=m$ and is 
   \be \label{olive}
  \begin{split} 
&
  \sum_{n=0}^{\infty} \frac{1}{(n!)^2}\  \ep_n \int  \bar J(x_1) \cdots \bar J(x_n)  \det \{ G(x_i,y_j) \}  J(y_1) \cdots J(y_n)  dx dy \\
  \end{split} 
 \ee
 If $\pi$ is a permutation of $(1,\dots, n)$ 
 then  $\sgn(\pi) = \sgn( \pi^{-1}) $ and  then 
  \be 
  \begin{split} 
  &  \int  \bar J(x_1) \cdots \bar J(x_n)  \det \{ G(x_i,y_j) \}  J(y_1) \cdots J(y_n)  dx dy \\
   = &   \int  \bar J(x_1) \cdots \bar J(x_n) \B[   \sum_\pi  \sgn \pi \prod_i G(x_i,y_{\pi(i)} )  \B]  J(y_1) \cdots J(y_n)  dx dy \\
     = & \sum_{\pi} \sgn \pi   \int  \bar J(x_1) \cdots \bar J(x_n)  \prod_i  G(x_i,y_i  )  J(y_{\pi^{-1} (1)} ) \cdots J(y_{\pi^{-1} (n) } )  dx dy \\
     = & \sum_{\pi}   \int  \bar J(x_1) \cdots \bar J(x_n)  \prod_i G(x_i,y_j  ) J(y_{1} ) \cdots J(y_{n } )  dx dy \\
        = & n! \ep_n < \bJ,  G  J >^n \\
 \end{split}
 \ee
 Substitute this in (\ref{olive}),
   use  $\ep_n^2 =1$, and get  $e^{ <\bar J, G J>} $.   $\blacksquare$
 \bigskip

 We now    assume the operator $G$ is translation invariant.    So it  is given by a convolution with a smooth function,  also denoted $G$.  
 We use the notation
 \be
 ( G f )_a (x) = \sum_b  \int G_{ab} (x-x') f_b(x') dx'    \hs   ( fG)_b(y) = \sum_a  \int f_a(y')G_{ab} (y'-y) dy' 
\ee
with the internal index still suppressed.   A norm on the function $G$  is defined by 
\be   \label{felix1} 
\| G\| = \max \left\{     \sup_a   \sum_b  \int|G_{ ab} (x)|  dx, \  \    \sup_b   \sum_a  \int |G_{ ab} (x)|  dx \right\} 
\ee
Then  $\| Gf\| _{\infty} \leq \|G\|   \|f \|_{ \infty } $ and  $ \|f G \| _{\infty} \leq  \|G\|  \|f \|_{ \infty } $.  But  integrating by parts we can transfer derivatives 
from $Gf$ to $f$  and so establish more generally 
\be    \label{felix2}  
    \| G f \|_{\cC^k}   \leq \|G\|   \| f \|_{\cC^k}   \hs       \|   f G\|_{\cC^k}   \leq \|G\|   \| f \|_{\cC^k}  
\ee
With these assumptions on $G$  we have the following specific estimates. 
 
 \begin{prop} \label{skunk}
\be  \label{skunk1} 
 \| F ( G J, \bJ G)  \|_h  \leq  \| F \|_{ \|G\|h}
 \ee 
\be \label{skunk2} 
 \| F (\psi +  G J, \bpsi  +\bJ G )  \|_{h, h'}  \leq  \| F \|_{ h+ \|G\|h'}
 \ee 
\be \label{skunk3} 
\| \int  F (\psi +   G J,\bpsi +  \bJ G )  d \mu_G( \psi )  \|_{h'}  \leq  \|F \|_{h(G) + \|G\| h'} 
 \ee
 \end{prop} 
 \bigskip
 
  \pr     $F'(J, \bJ ) = F ( G J, \bJ G ) $ has kernels
 \be
 \begin{split} 
& F'_{nm} ( x'_1, \dots, x'_n,y'_1, \dots, y'_m) \\=  
 &\int  F_{nm} ( x_1, \dots, x_n,y_1, \dots, y_m)    \prod_{i=1} ^n G(x_i-x_i') \prod_{j=1}^m G(y'_i - y_i) dx dy\\
 \end{split} 
 \ee
 where the pairing of spinor indices matches the pairing of spatial variables.
 Then for  $f \in \cC= \cC_k$   
  \be
 \begin{split} 
|<  F'_{nm}, f> |= & \left|  \blan  F_{nm}, (G \otimes  \cdots \otimes  G)  f \bran  \right| \\
\leq  & \| F_{nm} \|_{\cC'}  \|  (G \otimes  \cdots \otimes  G)  f  \|_{\cC}   \\
\leq  &  \| F_{nm} \|_{\cC'}\|G\|^{n+m}  \|f\|_{\cC} 
\end{split}
\ee
Hence 
\be 
\| F'_{nm} \|_{\cC'}  \leq  \|F_{nm}\|_{\cC'}  \| G \|^{n+m}   
 \ee
 which gives the first estimate (\ref{skunk1}).

 The second estimate (\ref{skunk2}) follows from the first and the fact that 
 if   $F^+ ( \psi , \psi' ) \equiv  F( \psi + \psi') $ then $ \| F^+ \|_{h,h'}  =   \| F\|_{h+h'}$ (Lemma 9 in\cite{DiYu24}).
 Indeed we have 
 \be
  F (\psi +  G J, \bpsi + \bJ G ) = F^+ ( \psi, \bpsi, GJ, \bJ G)  = (F^+ )' (  \psi, \bpsi, J,  \bJ ) 
 \ee
 and 
 \be  
\|(F^+ )' \|_{h,h'}  \leq \|F^+  \|_{h, \|G\| h' }   =  \| F \|_{h + \|G\| h' } 
\ee
 
 The third estimate (\ref{skunk3})  follows from 
 \be
 \| \int   (F^+ )' (  \psi, \bpsi, J, \bJ ) d \mu_G( \psi )  \|_{h'} 
 \leq  \|  (F^+ )'  \|_{h(G) , h'}  \leq    \| F \|_{h(G)  + \|G\| h' } 
 \ee
 \bigskip

 \begin{prop} \label{seven} 
  Let   $F \in \cG_h$ and $ h(G)  + \| G \| h' < h$.  Then in  $\cG_{h'} $ 
 \be \label{basic} 
 \int e^{ < \bar J, \psi>  + < \bpsi, J > } F\B( \psi, \bpsi\B )  d\mu_G (\psi) = e^{ <\bar J, G J>} \int  F\B( \psi + GJ, \bpsi +  \bJ G \B)  d\mu_G (\psi) 
 \ee 
 \bigskip
 \end{prop}

 \rems 
 \begin{enumerate}  
 \item
 This is the key identity for subsequent developments.   Formally    the result follows
 by the change of variables $\psi  \to \psi + GJ$ and $\bpsi \to  \bpsi + \bJ G $.   Indeed formally 
   $ d\mu_G (\psi)  = Z^{-1}   \exp ( - <\bpsi, G^{-1}  \psi>) D \psi $ so under the change it generates the object
   $ - < \bar J, \psi>  -  < \bpsi, J > - < \bar J, GJ >  $.   The existing  $< \bar J, \psi>  + < \bpsi, J >$ becomes
   $< \bar J, \psi>  + < \bpsi, J > + 2 < \bar J, GJ > $.  The two contributions add to $< \bar J, GJ > $. 
 
  \item 
 By  Proposition  \ref{skunk}  the bound 
  $ h(G)  + \| G \| h' < h$ ensures that the right side of the identity is well defined as an element of  $\cG_{h'} $ by (\ref{skunk3}).  The left side is well-defined 
  since the integrand is in $\cG_{h,h'} $ and the integral maps to $\cG_{h'}$  for     $h(G)< h $.  Thus we only have to show equality. 
 \end{enumerate} 
 \bigskip 
 
\pr (A.) 
We first suppose  $F = e^{ < \bpsi, K \psi> } $ for a smooth function $K(x,y)$.   Let $KG$ be the bounded operator
on    $L^2(\bbT) $ and assume  $\| KG \|<1$.  Let $Z = \int  e^{ < \bpsi, K \psi> }   d\mu_G (\psi) = \det ( I -KG ) $.  Then on the left we have
by Propositions \ref{three} and \ref{five} 
\be
\begin{split} 
 \int e^{ < \bar J, \psi>  + < \bpsi, J > } e^{ < \bpsi, K \psi> }   d\mu_G (\psi)
= &  Z  \int e^{ < \bar J, \psi>  + < \bpsi, J > } d\mu_{G,K }  (\psi)\\
= & Z \exp \B(  < \bJ,  G ( I -KG)^{-1} J>  \B)  \\
\end{split} 
\ee
On the right we have
\be \label{algae} 
\begin{split} 
&e^{ <\bar J, G J>} \int  \exp \B(< (\bpsi +\bJ G ), K ( \psi + GJ)>  \B)   d\mu_G (\psi) \\
= &   Z  e^{ <\bar J, G J>} \int  \exp \B(< \bJ, G K  GJ>  +< \bpsi , K  GJ> + <\bJ ,G K  \psi>  \B)  d\mu_{G,K }  (\psi) \\
=  & Z  e^{ <\bar J, ( G + G K  G )J>} \int  \exp \B( < \bpsi ,  J'> + < \bar J' ,  \psi>  \B)  d\mu_{G,K }  (\psi) \\
\end{split} 
\ee
where $J' = KGJ$ and $\bJ'  = \bJ GK$.
Now Proposition \ref{five} holds just as well with  $J' , \bJ'$ 
so the last integral in (\ref{algae}) is 
\be 
\begin{split} 
 \exp \B(   <\bar J', G(I-KG)^{-1}   J'     >    \B)  =   & < \bar J,G  KG(I-KG)^{-1} KG J     >   \\
=  & \exp \B( < \bar J,\B( -G -GKG +G (I-KG)^{-1}\B)   J     >   \B) \\
\end{split}   
\ee
Here  we used the identity
$A(I-A)^{-1} A = -I-A + (I-A)^{-1} $  with $A = KG$. 
The   first two terms cancel in (\ref{algae})  and we have again
$
 \sZ  \exp \B(  < \bar J,G (I-KG)^{-1}   J     >   \B)   
 $ 
 Thus the identity is established in this case.   
\bigskip

\noindent (B.) 
Now take  $K = K(s)$ of the form 
\be
K(s, x,y)  = \sum_{i=1}^n s_i  f_i(x) g_i (y ) 
\ee  
for arbitrary  smooth functions   $f_i,g_i$   and $s_i$ complex and sufficiently small depending on $f_i,g_j, G$.   Thus 
\be
< \bpsi, K(s)  \psi> = \sum_{i=1}^n s_i  \bpsi(f_i)  \psi (g_i)  
\ee
and the    result in part A says 
 \be
 \int e^{ < \bar J, \psi>  + < \bpsi, J > } e^{  < \bpsi, K(s)  \psi>} d\mu_G (\psi) 
 = e^{ <\bar J, G J>} \int    e^{  < (\bpsi + \bJ G ), K(s) (\psi +  GJ) >}            d\mu_G (\psi) 
 \ee 
  Taking derivatives  $\pa^n/\pa s_1 \cdots \pa s_n [ \cdots ]_{s_i =0} $ and rearranging the order on both sides yields 
 \be  \label{igor1} 
 \begin{split} 
& \int e^{ < \bar J, \psi>  + < \bpsi, J > }  \psi (f_1) \cdots \psi(f_n) \bpsi(g_1) \cdots\bpsi(g_n)   d\mu_G (\psi) \\
 & = e^{ <\bar J, G J>} \int   (\psi+GJ )(f_1) \cdots  (\psi+ GJ )(f_n) ( \bpsi + \bJ G) (g_1) \cdots   ( \bpsi +   \bJ G)(g_n)   d\mu_G (\psi)  \\ 
 \end{split} 
 \ee 
 The differentiation under the integral sign is justified since,  for example on the left,  the $\cG_{h,h'} $ valued function 
 $s_i \to  e^{ < \bar J, \psi>  + < \bpsi, J > } e^{  < \bpsi, K(s)  \psi>}$ is strongly  differentiable and  $F(\psi, J) \to\int F(\psi,J) d \mu_G(\psi)$
 is a continuous linear  function from $\cG_{h,h'} $ to $\cG_h$ as in Proposition \ref{four}.  

We now have established  that
 \be \label{igor2} 
 \int e^{ < \bar J, \psi>  + < \bpsi, J > }  F_{nn} (\psi, \bpsi)     d\mu_G (\psi)  = e^{ <\bar J, G J>} \int  F_{nn} \B(  \psi+ GJ,  \bpsi + \bJ G \B)   d\mu_G (\psi)   
 \ee 
 for  
 $
F_{nn}  = f_1 \otimes \cdots \otimes f_n \otimes   g_1 \otimes \cdots \otimes g_n    
 $
 with smooth $f_i, g_i$.  (Actually by our definition of the product it is this anti-symmetrized.  But since we are pairing it with something anti-symmetric we
 can take it as is.) 
 By a kernel theorem we can extend the result to any smooth  $F_{nn} $.  We show below that the identity can be extended  to $F_{nn} \in \cC'$.  Finally 
 we   multiply by $1/n!^2$ and sum over $n$.   We already know the sum converges and so   have the result  (\ref{basic}).  

 \bigskip
 
 \noindent (C.) 
 We  need the fact that for  any distribution $F$ on a torus,   there exists a sequence of smooth functions $F_N $ such that $F_N \to F$ weakly, i.e.
 $< F_N-F, f> \to  0$ for any smooth function $f$.  In Appendix B we give a   proof  of this  standard fact. 
 
  To use this weak convergence  note that  to establish the identity  it is sufficient to match the kernels of like powers of $J , \bJ$.  Once 
  the powers of $J, \bJ$ are fixed we know the identity holds with $F_{N, nn}$.  It suffices to show that on each side the expression with  $F_{N,nn}$
 converges to the expression with $F_{nn}$.  
 
 On the left side then consider 
 the term with $ < \bar J, \psi> ^r  < \bpsi, J >^r$ 
 which has the kernel in $J, \bJ$ proportional to 
 \be
\int  \B[  \int \psi( z) \bpsi(w )  F_{nn}(\psi, \bpsi )   \B]   d \mu_G 
\ee  
  where 
   $\psi( z)= \psi(z_1) \cdots \psi(z_r) $, $ \bpsi(w) = \bpsi(w_1)   \cdots \bpsi(w_n)$. 
We look at the difference with the regularization  which is
 \be
\int  \B(   F_{nn} -  F_{N,nn } \B)  (x,y )         \B[  \int \psi( z) \bpsi(w ) \psi(x)  \bpsi(y)   d \mu_G  \B]  \ dxdy
\ee  
This goes to zero since $G(x,y)$ is smooth and hence so is the bracketed expression which is a determinant in $G(x,y)$.

 On the right side  ignore the prefactor $e^{ <\bar J, G J>}$.  We consider a term   with the number of   $\psi $ and $\bpsi$ each equal to $p$,  and the number  of $J$ and $\bJ$ each equal to  $q$ with $p+q =n=m$.
 The kernel of such a term  in $J, \bJ$  is proportional to 
 \be
  \label{urgent} 
 \int F_{nn} ( x,y, x',y')      \B[     \int     \psi(x) \bpsi(y)   d \mu_G( \psi)   G( x' -z ) G(w -y')   \B]  dx dy dx'dy' 
   \ee
Here    $x = (x_1, \dots, x_p) $ and $x' =(x_1', \dots x_q')$ and $G(x') = G(x_1') \cdots G(x_q')$, etc. 
 Again the difference 
  \be 
 \int  \B(   F_{nn} -  F_{N,nn } \B) ( x,y, x',y')      \B[    \int     \psi(x) \bpsi(y)   d \mu_G( \psi)    G( x' -z ) G(w -y')   \B]  dx dy dx'dy' 
   \ee
converges to zero since the bracketed expression is smooth. $\blacksquare$

 
 \section{Correlation functions} \label{npoint} 

\subsection{generating functions}   \label{under} 
 Now we return to the two dimensional torus $\bbT_M$ with cutoff  covariance $G$ defined  by (\ref{fine}), and initial  action $S$ given by (\ref{initalS}) with renormalized parameters
$ g_0, z_0$, and with norms
 based on $\cC_k(\bbT_M) $ with $k=3$. 
We study correlation functions following a method   used in Dimock,Hurd \cite{DiHu92}, and Brydges, Dimock, Hurd \cite{BDH95} for scalar fields. 

 A generating  function  for correlations is
\be     \label{cheerios} 
\Om (J) =  \int     e^{ < \bar J, \psi>  + < \bpsi, J > }e^{ S( \psi)  }   d\mu_G (\psi) 
 \ee
   This is well-defined  in $\cG_{h'}$   for any $h'$ since   the integrand is in $\cG_{h(G),h'} $.  
   But   $h(G) $ and $\| S \|_{h(G)}$  are not bounded in
 the ultraviolet cutoff $N$, so this is not the end of the story.  
 As an element of $\cG_{h'} $ it has a representation 
 \be \label{cheerios2} 
 \Om (J) =\sum_{k,\ell }   \frac{1}{ k! \ell !}  \int_{  \bbT^k_M \times \bbT^{\ell}_M } 
 \Om_{ k \ell}( z_1, \dots, z_k,w_1, \dots,w_{\ell} )    J (z_1) \cdots J(z_n) \bJ(w_1) \cdots \bJ(w_{\ell} )    dz dw 
 \ee 
with kernels $\Om_{ k \ell}$ anti-symmetric in the $z$'s and $w$'s. 
 Only terms with $k=\ell$ are non-zero.

  The  normalized 2n-point function is 
 \be
 \blan \psi(x_1) \cdots \psi(x_n) \bpsi(y_1) \cdots  \bpsi(y_n)  \bran
=\sZ^{-1}   \int   \psi(x_1) \cdots \psi(x_n) \bpsi(y_1) \cdots  \bpsi(y_n) e^{ S( \psi ) } d \mu_G ( \psi) 
\ee
where now   $\sZ = \Om(0)$ is the partition function.  
These can be obtained from the generating function using as in (\ref{49}) 
\be
\begin{split} 
\frac{\pa}{ \pa \bJ(x) } e^{<\bJ, \psi> } = & e^{<\bJ, \psi> } \psi(x) \hs 
\frac{\pa}{ \pa J(y) } e^{<\bpsi, J > } =   e^{<\bpsi, J > }  (- \bpsi(y) ) \\
\end{split} 
\ee
This  yields 
\be
\begin{split} 
&  \blan \psi(x_1) \cdots \psi(x_n) \bpsi(y_1) \cdots  \bpsi(y_n)  \bran \\
  =& \Om(0) ^{-1} \left[\frac { \pa ^{2n} \Om(J) }{\pa \bJ(x_1) \cdots  \pa \bJ(x_n) \pa J(y_1) \cdots \pa J(y_n) } \right]_{J=0}  (-1)^n \\
  =&   \Om(0) ^{-1}  \Om_{nn}  ( y_1  \cdots y_n, x_1, \cdots x_n)   (-1)^n \\
    \end{split} 
  \ee

Now suppose we can find $\La(J) = \La(J, \bJ)$ so that  $\Om(J )  = e^{\La(J ) } $.   Then
$\La(J) $ generates the truncated $2n$ point functions by  
 \be \label{donkey} 
\begin{split} 
   \blan \psi(x_1) \cdots \psi(x_n) \bpsi(y_1) \cdots  \bpsi(y_n)  \bran ^T 
 = & 
  \left[\frac { \pa ^{2n} \La  (J) }{\pa \bJ(x_1) \cdots  \pa \bJ(x_n) \pa J(y_1) \cdots \pa J(y_n) } \right]_{J=0}  (-1)^n \\
 =&  \La_{nn}  ( y_1  \cdots y_n, x_1, \cdots x_n) (-1)^n  \\
   \end{split} 
  \ee
 These can be expressed in terms of the ordinary $2n$-point functions.   
Here are some examples.    
\be 
0=   \blan \psi(x) \bran =    e^{-\La(0) }    \left[\frac {\pa e^{\La(J)}    }{\pa \bJ(x)   }      \right]_{J=0}    =   \blan \psi(x) \bran^T 
\ee
and
 \be
\begin{split} 
  \blan \psi(x) \bpsi (y)   \bran    
   =  & e^{-\La(0) }    \left[\frac {\pa e^{\La(J)}    }{\pa \bJ(x)  \pa J(y) }      \right]_{J=0} (-1)  \\
   =&  \left[    \frac{ \pa^2 \La(J)  } {\pa \bJ(x)   \pa J(y) }  -   \frac{ \pa \La(J)  } {\pa \bJ(x)   }  \frac{ \pa \La(J)  } {  \pa J(y) }        \right]_{J=0}  (-1) \\
   = &   \blan \psi(x) \bpsi (y)   \bran^T
 \end{split}   
 \ee
 Similarly 
 \be
 \begin{split} 
  & \blan \psi(x_1) \psi(x_2) \bpsi (y_1) \bpsi(y_2)  \bran   
 =  \blan \psi(x_1) \psi(x_2) \bpsi (y_1) \bpsi(y_2)  \bran^T \\
 & -    \blan \psi(x_1) \bpsi (y_1)   \bran  \blan  \psi(x_2)  \bpsi(y_2)  \bran  +  \blan \psi(x_1) \bpsi (y_2)   \bran  \blan  \psi(x_2)  \bpsi(y_1)  \bran
\end{split} 
 \ee
 gives an expression for   $ \blan \psi(x_1) \psi(x_2) \bpsi (y_1) \bpsi(y_2)  \bran^T $, and so forth. 
\bigskip

 For this to be useful we need   a workable expression for  $\La(J) $ with good bounds.  This is just what is supplied by our renormalization group
 analysis  as we now explain. 
 Starting with the expression (\ref{cheerios}) 
  we scale up  from  $\bbT_M$ to  $\bbT_{M+N} $ and find 
\be
\Om(J)  =  \int     e^{ < \bar J_0, \psi>  + < \bpsi, J_0 > }e^{ S_0( \psi)  }   d\mu_{G_0}  (\psi) 
\ee 
 where  
 $
 J_0 (x ) = J_ {L^N}(x)  \equiv L^{-\frac32N} J(L^{-N} x)
 $
 is defined so  
 $
  < \bJ, \psi_{L^{-N}} >  =  < J_0, \psi >
 $. 
 More generally we define on $\bbT_{M+N-k} $ 
  \be
 J_k (x ) = J_ {L^{N-k} }(x)  \equiv L^{-\frac32(N-k) } J(L^{-(N-k) } x)
 \ee
 Recall also that $w_k$ is defined by  $w_{k+1} = (w_k + C_k)_{L^{-1} }$  
and is given explicitly in (\ref{wk}).

  \begin{lem} 
 For any  $0 \leq k \leq N$
 \be 
\Om(J ) 
 =     e^{ <\bar J_k, w_k J_k>}    \int   e^{ < \bar J_k, \psi>  + < \bpsi, J_k > } \exp \B(  S_k  (\psi +  w_k J_k  , \bpsi  +  \bJ_k w_k  )   \B )  d \mu_{G_k} (\psi) 
\ee
 \end{lem}

 \pr  We have  $w_0 =0$ so the assertion holds for $k=0$.  We assume it holds for  $k$ and prove it for $k+1$. 
 The following is modification of  (\ref{start}) - (\ref{stop}).  
 We first  split the integral into a fluctuation integral and an background integral
 and write
  \be  \label{lolly} 
 \begin{split} 
&\Om(J) =   e^{ <\bar J_k, w_k J_k>}   \int   e^{ < \bar J_k, \psi >  + < \bpsi  , J_k > }  [ \cdots ]  d \mu_{G_{k+1,L}} ( \psi)  \\ 
   \end{split} 
 \ee
where the bracketed expression is.   
\be
 [ \cdots ]  =  \int   e^{ < \bar J_k, \eta >  + <  \bar \eta , J_k > }  \exp \B(  S_k  (\psi+ \eta + w_kJ_k  , \bpsi + \bar \eta    +    \bJ_k w_k )    \B )d \mu_{C_k} ( \eta  ) 
\ee
We apply proposition \ref{seven} to this  which gives
\be
\begin{split} 
 [ \cdots ]  =   &    e^{ <\bar J_k,  C_k   J_k>} \int 
 \exp \B(  \ S_k  \big(\psi +\eta +  ( w_k+ C_k) J_k  , \bpsi    + \eta +    \bJ_k( w_k + C_k)   \big )   \B ) d \mu_{C_k} ( \eta  )  \\
 =   &    e^{ <\bar J_k,  C_k   J_k>} 
\exp \B(  \tilde S_{k+1}  \big(\psi + ( w_k+ C_k) J_k  , \bpsi    +   \bJ_k( w_k + C_k)    \big)   \B ) \\
 \end{split} 
 \ee
Insert this back in (\ref{lolly}) and find
 \be  \label{lolly2} 
 \begin{split} 
&\Om(J) =   e^{ <\bar J_k, (w_k+ C_k)  J_k>} \\
&  \int   e^{ < \bar J_k, \psi >  + < \bpsi  , J_k > }  \exp \B(  \tilde S_{k+1}  \big (\psi + ( w_k+ C_k) J_k  , \bpsi    +   \bJ_k( w_k + C_k)   \big)    \B )  d \mu_{G_{k+1,L}} ( \psi)  \\ 
   \end{split} 
 \ee
Finally scale down from $\bbT_{M+N -k} $ to $\bbT_{M+N -k-1} $.   Since $w_k + C_k$ scales to $w_{k+1} $ and $J_k$ scales to $J _{k+1} $ we get 
this desired
 \be  \label{lolly3} 
 \begin{split} 
&\Om(J) =   e^{ <\bar J_{k+1} , w_{k+1}  J_{k+1} >} \\
&  \int   e^{ < \bar J_{k+1} , \psi >  + < \bpsi  , J_{k+1}  > }   \exp \B(  S_{k+1}   (\psi +  w_{k+1}  J_{k+1}   , \bpsi    +   \bJ_{k+1}  w_{k+1}   )      \B )  d \mu_{G_{k+1}} ( \psi)  \\ 
   \end{split} 
 \ee
 
The application  Proposition \ref{seven}  was justified provided   $h$ is large enough that 
\be
h(C_k)  + \|C_k \| h'  \leq h
\ee
Here  $h$ is already  constrained by our basic theorem which included the condition    $h(C_k) < h$.    The $h'$ is arbitrary and  $C_k(x) = \cO( e^{-|x|/L }) $ \cite{DiYu24} ,
 so  $\|C_k\| \leq \cO(L^2)  $  which can also be beat by large enough $h$.     
 $\blacksquare$
 \bigskip
 
 Now we make a final adjustment: 
 
 \begin{lem} 
  \be \label{stir} 
\Om(J ) 
 =     e^{ <\bar J, G J>}    \int  \exp \B(  S_N  (\psi + G J  , \bpsi  +  \bJ G  )   \B )  d \mu_{G_N } (\psi) 
\ee
\end{lem} 
 
 \pr 
Consider the final  expression for $\Om(J) $  on $\bbT_M$.  Since $J_N = J$ this is 
  \be 
\Om(J ) 
 =     e^{ <\bar J, w_N J>}    \int     e^{ < \bar J, \psi>  + < \bpsi, J > }\exp \B(  S_N  (\psi +  w_N J  , \bpsi  +  \bJ w_N  )    \B )  d \mu_{G_N } (\psi) 
\ee
 we make a final application of Proposition  \ref{seven}     and this becomes
   \be 
\Om(J ) 
 =     e^{ <\bar J, (w_N+ G_N )  J>}    \int  \exp \B(  S_N  (\psi +  (w_N+ G_N )  J  , \bpsi  +  \bJ (w_N +G_N)   )   \B )  d \mu_{G_N } (\psi) 
\ee
Since   $w_N + G_N = G$  this is the announced result.

For the application   Proposition  \ref{seven}  we need
\be \label{inc} 
h(G_N)  + \|G_N  \| h'  \leq h
\ee
We claim   $h(G_N) $ (or any $h(G_k)$)  is  bounded by $\cO(1)$.     This is the case as in the discussion of $h(C_k)$ in  lemma 7 in \cite{DiYu24}.
 The key observation is that   $\tilde G_N$ on $\bbT^*_M$ can be split as
  \be
  \tilde G_N(p)  =  \frac{-i\slp}{p^2}   e^{- p^2 }  =  \left(  \frac{-i\slp}{|p|^{\frac32} }e^{-\frac12 p^2}  \right)   \left(  \frac{1} { |p|^{\frac12} }e^{-\frac12p^2}\right) 
    \equiv  \tilde  G_1(p)  \tilde G_2(p) 
  \ee
and both $ \tilde  G_1,   \tilde G_2$ are square summable, also with extra positive powers of $p$.    On the other hand $G_N(x-y) $
is actually independent of $N$ and is bounded   so  $\| G_N  \| \leq \one \Vol(\bbT_M) $.  Thus (\ref{inc})  is an allowed constraint on $h$, although not uniform in $M$. $\blacksquare$
\bigskip

Now we continue with an analysis of (\ref{stir}). 
With $\vep_N =0$   the final value of the action  from (\ref{bongo}) is 
\be \label{bongo2} 
\begin{split}
S_N(\psi)   = &     - z_N \int  \bpsi \slpa \psi   +  g_N \int  (\bpsi \psi )^2    +  Q_N^{\reg}(\psi)    +  E_N (\psi)   \\
& +   p_N \int ( \bpsi \ga_5 \psi )^2 +  v_N \int  \sum_{\mu}    (\bpsi \ga_{\mu} \psi )^2  + {Q_N ^{\reg}}' (\psi)  \\
\end{split} 
\ee
Define 
\be
  \Phi_N(\psi, \bpsi)  =        \int  \B(    e^{S_N(\psi' + \psi, \bpsi' + \bpsi)  } -1 \B)  d \mu_{G_N} (\psi') 
\ee
and then 
\be   \label{omj}
\Om(J )  =  e^{ <\bar J_, G J>}   ( 1 + \Phi_N(GJ ,\bJ G ) )      
\ee
This suggests that  for  $\Om(J) = e^{\La(J) }$ we should take
\be  \label{gen}  
\La(J)   =  <\bar J_, G J> + \log  \B ( 1 + \Phi_N(GJ ,\bJ G ) \B)    
\ee
To justify this we start with an analysis of 
 $\Phi_N( \psi, \bpsi) $.

 \begin{lem} \label{citrus}  Assume  the hypotheses of Theorems \ref{easter1},  \ref{easter2}  now with $g_f = g_N$ 
 sufficiently small  depending also on $\Vol(\bbT_M)$.
 \begin{enumerate}
 \item     There is a constant $C_{\Phi} $ (depending on $L, h, C_Z$)   such that  
 \be
 \begin{split} 
   \|S_N \|_h     \leq  &  \frac12 C_{\Phi}  \Vol(\bbT_M)  g_f  \\
 \| \Phi_N \|_{\frac12 h}  \leq &  C_{\Phi} \Vol ( \bbT_M) g_f  \\
 \end{split} 
 \ee
 \item For $\Phi_N =  \Phi_N(\psi, \bpsi)$,     $\log ( 1 + \Phi_N  ) $ is  well-defined as power series in $\Phi_N$, it exponentiates to  $ 1 + \Phi_N  $,  
 and satisfies the bound
  \be \label{goat} 
 \|  \log ( 1  + \Phi_N  ) \|_{\frac12 h}           \leq 2 C_{\Phi} \Vol ( \bbT_M) g_f                 
 \ee
\end{enumerate} 
\end{lem} 
  \bigskip

 \rem  The partition function is    $\sZ =1 + \Phi_N(0) $ and $\Phi_N(0) \leq  \| \Phi_N \|_{\frac12 h} $.    So again we have the ultraviolet  stability bound 
 \be
 |\sZ - 1| \leq   C_{\Phi} \Vol ( \bbT_M) g_f 
\ee
This extends the result in \cite{DiYu24} where $\Vol ( \bbT_M)=1$ was assumed. 
\bigskip

 \pr  We have
 \be
 \begin{split} 
 \| g_f \int_{\bbT_M}  (\bpsi \psi)^2 \|_h \leq  &\frac14   h^4 \Vol(\bbT_M)g_f  \\
 \| z_N  \int_{\bbT_M}   \bpsi \slpa \psi  \|_h  \leq  & h^2 \Vol(\bbT_M)|z_N| \leq C_Z h^2  \Vol(\bbT_M)g_f  \\
 \end{split}
 \ee
 From lemma 12 in \cite{DiYu24},  $\| Q^{\reg}_N  \|_{h, \Ga_4}  \leq Ch^6g_f^2 $ and this yields  
  \be
 \begin{split}
   \|Q^{\reg}_N  \|_h   \leq &  \sum_X \|Q^{\reg}_N  (X) \|_h \leq  \sum_{\sq \subset \bbT_M}  \left( \sum_{X \supset \sq}  \|Q^{\reg}_N  (X) \|_h \Ga_4(X) \right) \\
  = & \Vol(\bbT_M)  \|Q^{\reg}_N  \|_{h, \Ga_4} \leq      Ch^6    \Vol(\bbT_M)  g^2_f \\
  \end{split} 
 \ee
 The  bounds on  $ \| E_N  \|_h,     \| {Q_N ^{\reg}}' \|_h $ are even smaller.  These combine to give the stated bound on $ \|S_N \|_h   $.

 We  can  assume  $C_{\Phi} \Vol(\bbT_M) g_f<1 $   and then $\| S_N\|_h < \frac12$ and 
 \be
 \|e^{S_N}  -1 \|_h  \leq  2\|S_N\|_h \leq C_{\Phi} \Vol(\bbT_M)  g_f
 \ee
 Let $S_N^+ ( \psi', \bpsi', \psi, \bpsi ) =   S_N( \psi' + \psi, \bpsi' + \bpsi) $.
 Then 
 \be
\| \Phi_N  \|_{\frac12h}  \leq  \| e^{S^+_N} - 1  \|_{ h(G_N) ,\frac12h } \leq \| e^{S_N} - 1  \|_{ h(G_N) + \frac12 h  }   \leq  \| e^{S_N} - 1  \|_h  \leq  C_{\Phi} \Vol(\bbT_M)  g_f
  \ee
 Here we have assumed   $h(G_N) < \frac12 h$ which is allowed as in (\ref{inc}).    Thus the bound on $\Phi_N$ is established. 

The second point follows  from the  discussion in section \ref{derivatives}.  $\blacksquare$
\bigskip 

\rem   We want to extend this analysis of $\Phi_N(\psi, \bpsi)$ to $\Phi_N(GJ ,\bJ G )$.    In this case we have  as in Proposition \ref{skunk} 
\be
 \| \Phi_N(G(\cdot)  ,(\cdot)G)  \|_{\frac12h}  \leq   \| \Phi_N \|_{\frac12h \| G\| } 
\ee
 For a bound on the   $L^1$ norm  $\| G \| $ we use $G = w_N  + G_N $.  
The     $w_N(x-y)$ has the summable  short distance singularity  $ \cO( |x-y|^{-1} ) $   and is exponentially decaying (lemma 27 in \cite{DiYu24}). 
so   $\|w_N \| \leq \cO( 1)  $.   We already noted  $\|G_N \| \leq  \cO(1) \Vol(\bbT_M) $.   Hence  $\|G\| \leq  \cO(1) \Vol(\bbT_M) $. 
   
 So if $g_f$ sufficiently small (now depending on $M$) 
the whole construction works with $h$ replaced by $\|G\| h$ and 
   $ \| \Phi_N(G(\cdot)  ,(\cdot)G)  \|_{\frac12h}  <1$.  Then again  $\log ( 1 + \Phi_N  ) $ is  defined and   exponentiates to  $ 1 + \Phi_N  $.  Hence $\Om(J) = e^{\La(J) }$ with  $\La(J) $ given by  (\ref{gen}).

  \bigskip

 \subsection{correlation functions} 
 We  consider the truncated functions as distributions introducing $\psi(f) = \int \psi(x) f(x)  dx$ and $\bpsi(g) = \int \bpsi(x) g(x) dx$. 
 We assume the hypotheses of lemma  \ref{citrus}.  
 The main result is the following: 
\bigskip

\begin{thm} 
There are constants $C_{2n}  $ such that for  $f, g \in \cC^3(\bbT_M) $
\be
|  \blan \psi(f)  \bpsi(g)  \bran ^T  |   \leq     C_2 \ \|f\|_{\cC^3} \|g\|_{\cC^3}  
\ee
and for $n \geq 2$ and $ f_i, g_i  \in \cC^3(\bbT_M) $
\be
|  \blan \psi(f_1)  \cdots  \psi(f_n)  \bpsi(g_1)  \cdots \bpsi(g_n)  \bran ^T |  \leq    C_{2n}  \ g_f  \prod_{i=1}^n \|f_i\|_{\cC^3} \|g_i\|_{\cC^3}  
\ee
\end{thm}

\rems
\begin{enumerate}
\item  The constants $C_{2n}$  (stated more precisely in the proof)   depend on  $n, C_{\Phi},\Vol(\bbT_M)$,
but not on the UV cutoff $N$ nor the coupling constants or  fields strength.   
\item   As is well-known the basic  correlation functions without truncation can be expressed in terms of the truncated correlation functions,  
and so  bounds of this form  hold for these  as well.
\end{enumerate} 
\bigskip

\pr  For the two point function we have  from (\ref{donkey}) 
\be  \label{inkspot} 
\begin{split} 
  \blan \psi(x)  \bpsi(y)  \bran ^T    = & -  \La_{11}  (y,x)   \\ 
  =  & G(x-y)  - \B[\log ( 1 + \Phi_N)\B(G(\cdot)  , (\cdot)  G \B)  \B]_{11} (y,x)  \\
  =&   G(x-y)  -  \int\B[\log ( 1 + \Phi_N)  \B]_{11} (y',x')  G(y'-y)  G(x-x')  dx'dy' \\
  \end{split} 
  \ee
   Also for  $n \geq 2$
  \be
  \begin{split} 
 &  \blan \psi(x_1)  \cdots  \psi(x_n)  \bpsi(y_1)  \cdots \bpsi(y _n)  \bran ^T =  \La_{nn}  ( y_1, \dots y_n, x_1, \dots x_n)  (-1)^n \\
     = &\int  \B[\log ( 1 + \Phi_N)  \B]_{nn}   (y'_1, \dots y'_n, x'_1, \dots x'_n)
 \prod_{i=1}^n G(y_i'-y_i )  G(x_i-x'_i)   dx' dy' (-1)^n  \\
 \end{split}
\ee

The  $ [  \log ( 1 + \Phi_N  ) ]_{nn} ( y'_1, \dots y'_n, x'_1, \dots x'_n) $ is a  distribution in $\cC'$ and  needs  test functions in  $\cC$, in other words  we need
 $ G(x - \cdot)  $ and  $G(\cdot - y ) $ to be in $\cC^3(\bbT_M) $.    This is true which is why this expression is well-defined.  
 But the $\cC^3 $ norms are not uniform in the ultraviolet cutoff, which is why we  treat the fields as distributions. 
As distributions we have 
\footnote{In general for $F_{nm} \in \cC'(\bbT^n \times \bbT^m) $ the multilinear functional  $F_{nm} (f_1, \dots, f_n, g_1, \dots, g_m) $ on $\cC^3(\bbT) $ is
defined as  $F_{nm} (f_1 \otimes \cdots  \otimes  f_n \otimes  g_1\otimes  \cdots \otimes g_m)$}
\be  \label{inkspot2} 
  \blan \psi(f)  \bpsi(g)  \bran ^T    = <f,  G g>   -  \B[  \log ( 1 + \Phi_N )  \B]_{11} ( Gg, fG)   
  \ee
and for $n \geq 2 $
\be
   \blan \psi(f_1)  \cdots  \psi(f_n)  \bpsi(g_1)  \cdots \bpsi(g _n)  \bran ^T 
    =  \B[  \log ( 1 + \Phi_N  )  \B]_{nn} (Gg_1, \dots, G g_n,    f_1G, \dots f_nG )(-1)^n 
\ee
Note that   $fG, Gg$   do have uniform   $\cC^3(\bbT_M) $ bounds.    Indeed from (\ref{felix2}) we have
\be \label{banana} 
\| fG \|_{\cC^3}   \leq \|G\|   \| f\|_{\cC^3}    \hs        \| G g \|_{\cC^3}   \leq \|G\|   \| g\|_{\cC^3} 
 \ee
 and as noted  $\|G\|$ does have a uniform bound. \bigskip 

The estimate (\ref{goat})  implies 
 \be
    \|  \B[  \log ( 1 + \Phi_N  )  \B]_{nn} \|_{\cC'}   \leq    (n!)^2 (2 h^{-1})^{2n}  \|  \log ( 1 + \Phi_N) \|_{\frac12 h}   \leq    (n!)^2   (2 h^{-1})^{2n}  2 C_{\Phi} \Vol(\bbT_M)  g_f
\ee
 and so   using  (\ref{banana}) 
  \be
 \begin{split}   
  &   \B[  \log ( 1 + \Phi_N  )  \B]_{nn} (f_1G, \dots f_nG, Gg_1, \dots, G g_n) \\
\leq   &  \|\   [ \log ( 1 + \Phi_N  )  ]_{nn}\|_{\cC'} \|G\|^{2n}   \prod_{i=1}^n  \|f_i \|_{\cC^3}  \|g_i \|_{\cC^3} \\
\leq &  \B[  (n!)^2\ (2\|G\| h^{-1})^{2n}  2  C_{\Phi} \Vol(\bbT_M)  \B]  g_f      \prod_{i=1}^n  \|f_i \|_{\cC^3}  \|g_i \|_{\cC^3} \\
\end{split}
\ee
This settles things for  $n \geq 2$    in which case the bracketed expression is $C_{2n}$.    For the two point function  we   have for the lead term
 \be
 | < f,   G g > |  \leq   \|G \|   \Vol(\bbT_M)    \|f \|_{\infty}  \|g \|_{\infty } 
      \leq      \|G \|     \Vol(\bbT_M)  \|f \|_{\cC^3}  \| g \|_{\cC^3} 
\ee
and in this case we take $C_2 = 2 \|G \|     \Vol(\bbT_M) $ since the other term is down by $\cO(g_f)$. 
$\blacksquare$

\subsection{first order corrections} 

Suppose we want to compute the $\cO(g_f)$ corrections to the correlation functions.   
We can proceed as follows.  Write  $\Phi_N(\psi)  = \Phi_N ( \psi, \bpsi)$ as $\Phi_N(1, \psi) $
where
\be
\Phi_N(t, \psi)  = \int  \B(   e^{tS_N(\psi' + \psi) } -1 \B) d \mu_{G_N} (\psi') 
\ee
This is an entire function of   $t$, but we consider $|t| \leq r$ with  $r = (C_{\Phi} \Vol(\bbT_M)  g_f)^{-1} $ in which case  $\|tS_N \|_h \leq \frac12$ and 
$ \| \Phi_N(t) ) \|_{\frac12h}  \leq 1$ by lemma \ref{citrus}.   
We  study  
 $\Phi_N(\psi)= \Phi_N(1, \psi )$ and  with a short expansion around $t=0$.  Since $\Phi_N(0) = 0$ this 
 has the form  
 \be
  \Phi_N(\psi) =  \Phi'_N(0, \psi) +   \frac{1}{2\pi i}  \int_{  |t| = r}  \frac{dt} { t^2(t-1) }     \Phi_N(t, \psi)  \equiv  \Phi'_N(0, \psi) +  \Phi^+ _N( \psi)
\ee
The remainder satisfies $ \| \Phi^+_N\|_{\frac12h}  \leq r^{-2} = \cO(g_f^2)$, so to first order we only need to study $ \Phi'_N(0, \psi) $ which is
 \be 
  \Phi'_N (0, \psi)  =  \int   S_N(\psi' + \psi)     d \mu_{G_N} (\psi') 
  \ee
  Inserting the expression for $S_N$  from (\ref{bongo2})  we find  with $z_f = z_N, g_f = g_N$
  \be
  \begin{split} 
   \Phi_N'(0, \psi)  = & \int   \left(  - z_f \int  (\bpsi' + \bpsi)\slpa (\psi' + \psi)   +  g_f \int \B( ( \bpsi' + \bpsi) (\psi' + \psi )  \B)^2\right)     d \mu_{G_N} (\psi') \\
 + &  \cO(g_f^2) \\
 \end{split} 
\ee
Since  $ \int \psi'(x)  \bpsi'(x)    d \mu_{G_N} (\psi')  =   G_N(0)  =0$ this simplifies to  
  \be
   \Phi_N'(0, \psi)  =     - z_f \int  \bpsi \slpa \psi    +  g_f \int   (  \bpsi \psi )^2   
 +   \cO(g_f^2) + \const 
\ee
For the generating function
\be \label{gen2}  
  \begin{split} 
   \Phi_N'(0, GJ, \bJ G )  = &      - z_f  \int \bJ G\slpa GJ     +  g_f \int   \B( (\bJ G)( G J)   \B)^2
 +   \cO(g_f^2) + \const \\
 \end{split} 
\ee

Now for the two point function instead of (\ref{inkspot2}) 
we have
\be   
\begin{split}
  \blan \psi(f)  \bpsi(g)  \bran ^T    =& <f,  G g>   -  \B[  \log ( I + \Phi_N )  \B]_{11} ( Gg,fG)   \\
= &    <f,  G g>   -   [ \Phi_N ]_{11} ( Gg,fG)  + \cO(g_f^2) \\
= &    <f,  G g>   -   [ \Phi'_N(0)  ]_{11} ( Gg,fG)  + \cO(g_f^2) \\
 \end{split} 
  \ee
 And from   (\ref{gen2}) this is evaluated as 
 \be  \label{inkspot3} 
\begin{split}
  \blan \psi(f)  \bpsi(g)  \bran ^T  
= &     <f,  G g>     -z_f  <f,   G \slpa G g>   + \cO(g_f^2)    \\
 \end{split} 
  \ee
  
Similarly for the   four point function
\be   \label{inkspot4} 
\begin{split} 
   \blan \psi(f_1)   \psi(f_2)  \bpsi(g_1)\bpsi(g _2)  \bran ^T 
     = &[  \Phi'_N(0)  ]_{22} (Gg_1,   Gg_2, f_1G, f_2G)   + \cO(g_f^2)\\
\end{split}
\ee
This is evaluated from   (\ref{gen2}) as
\be   \label{inkspot5} 
\begin{split} 
   \blan \psi(f_1)   \psi(f_2)  \bpsi(g_1)\bpsi(g _2)  \bran ^T 
     = &g_f \int  \B( (f_1G) (z)( Gg_1)(z) \B)\B ( (f_2G)(z)( Gg_2)(z) \B)  dz  \\
     - &g_f \int   \B ((f_1G)(z)( Gg_2)(z) \B) \B( (f_2G) (z)( Gg_1)(z) \B) dz   + \cO(g_f^2)\\
\end{split}
\ee
For the other truncated correlation functions there is no $\cO(g_f)$ contribution. 

\subsection{concluding remarks} 
 \begin{enumerate} 
 \item   Similarly one can explicitly compute   corrections  second order  in the coupling constant, taking advantage of the fact
 that in the effective actions we  have an explicit expression for  the second order contribution in  $Q^{\reg}_N$. 
   To go to a higher order one would need to carry higher order terms  in the effective actions. 
    
 \item The first two terms in (\ref{inkspot3})  can be written pointwise as
\be  \label{ground} 
 G(x-y)   - z_f  (G \ \slpa G)(x-y)  =    G_{z_f} (x-y)   + \cO(g_f^2) 
\ee
where
\be 
 G_{z_f}  (x-y) =   L^{-2M }  \sum_{ p  \in  \bbT_M^*, p \neq 0}  e^{ip(x-y)}  \frac{-i\slp}{p^2} \frac{1}  {z_f + e^{   p^2/L^{2N} } }
\ee
As $N \to \infty$  this  converges to  the  renormalized  Dirac propagator with no cutoffs 
\be
(1+z_f)^{-1}   L^{-2M }  \sum_{ p  \in  \bbT_M^*, p \neq 0}  e^{ip(x-y)}   \frac{-i\slp} {p^2}   
\ee  
To first order this  what we would expect from inspection of the effective action.

\item  
One can undoubtedly  show that to first order  the short distance singularity of the two point function is $\cO(|x-y|^{ -1} )$.
For example in (\ref{ground})   $( \slpa G) (x-y) $ is an approximate  delta function, and we know that $G(x-y)  = \cO(|x-y|^{ -1} )$ from the remark at the end
of section \ref{under}.   

 But one would really
like to know that the short distance behavior of   the full two point function is  $\cO(|x-y|^{ -1} )$. 
   This may require consideration of the whole history of the 
effective actions $S_0, S_1, S_2 , \dots$ rather than just the final  $S_N = S_f $.  This was found to be  the case by  Gawedzki, Kupiainen 
who had a similar problem 
in their treatment of infrared problems for certain scalar field models \cite{GaKu84}, \cite{GaKu86}. 
\end{enumerate} 


\newpage

\begin{appendix}

\section{The weight function $\Ga(X) $} \label{A} 

The weight   factor $\Ga(X)$  is the same as that used in \cite{BDH95} - \cite{DiHu92}. We give the definition in dimension $d$.   It is defined by 
 \be \label{Gamma} 
  \Ga(X) = A^{|X| } \Theta(X) 
  \ee
 Here  $|X| $ is the number of  unit  squares in  the paved set  $X$, so $|X|$ is the volume of $X$.
In $A^{|X|} $ the $A$ is a sufficiently large constant depending on $L$,  say $A = L^{d+2}$.  Finally 
  $\Theta(X)$ has the form
  \be
  \Theta(X)  = \inf_T \prod_{\ell \in T } \theta(| \ell | )  
  \ee 
 where the infimum  is over all tree graphs  $T$ on the centers of the blocks in $X$.  The  $\ell$ are the lines in the graph 
 and   $|\ell |$ is the length in an $\ell^{\infty}$ norm.   The function $\theta(s)$ is defined by  $\theta(0) = \theta(1) =1$ and for $s=2,3, \dots $ 
\be 
\theta(s)  =  (2L^{d+1} ) ^n \hs  \textrm{  if } \hs  L^n <  s \leq L^{n+1} 
\ee
Then if   $L^n < s \leq  L^{n+1} $
we have   $L^{n-1}  <  s/L  \leq  L^{n} $ and so $L^{n-1}  < \{ s/L \}  \leq L^{n} $   where $\{ x\} $ is the smallest integer greater than or equal to $x$. 
Therefore $\theta$ has the useful scaling behavior
 \be
 \theta (   \{ s/L \}  ) = (2L^{d+1} ) ^{n-1}    =  (2L^{d+1} ) ^{-1} \theta(s) 
\ee
Note also  that  for $ L^n <  s \leq L^{n+1}  $
\be
\theta(s)  =   (2L^{d+1} ) ^n   =   2^n    ( L^n ) ^{d+1}  \leq  2^n   s^{d+1}   \leq  L^n   s^{d+1} \leq s^{d+2} 
\ee
Hence the bound holds for    $s=1,2, 3\dots$.

 We also will use  a modification defining for positive integer $n$  
 \be
 \Ga_n(X)  = e^{n|X|} \Ga(X) 
 \ee

\newpage

\section{Smooth approximations to distributions} \label{B} 

Let $f$ be a smooth function on a torus $\bbT$.  Define Fourier coefficients for $p \in \bbT^*$  by 
\be
\tilde f(p)  = <f, e_{-p}  >  \hs \hs e_p(x)  = e^{ipx} 
\ee
Then $\tilde f(p)$ is rapidly decreasing in $p$ and
it is a standard fact that the partial sum of the  Fourier series
\be
f_N(x)  \equiv   
 \Vol(\bbT) ^{-1} \sum_{|p| \leq N}  \tilde f(p) e_p(x) 
 \ee
converges uniformly to $f(x)$,  also for any number of derivatives. 

Now let  $F$ be a distribution on $\bbT$.  Then $F$ is continuous on some $\cC^k(\bbT)$, and  there are constants $C,k$ such that 
\be
| < F, f> | \leq C  \sup_{|\al | \leq  k}  \| \pa ^{\al} f  \|_{\infty}  
\ee
  Fourier coefficients defined  by  $\tilde F(p)  = <F, e_{-p}  > $ satisfy  
 \be
 |\tilde F(p) |  \leq  C  \sup_{|\al | \leq k }  \| \pa ^{\al} e_p  \|_{\infty}   \leq C'  ( 1 + p^2 ) ^{k/2}
\ee 
Then 
\be
<F, f_N> =  \Vol(\bbT) ^{-1}  \sum_{|p| \leq N } < F, e_p> \tilde f(p)  =  \Vol(\bbT) ^{-1} \sum_{|p| \leq N } \tilde F(-p)  \tilde f(p)  
\ee
By  the uniform convergence of $f_N$ and its derivatives,   and the fact that  $ \tilde F(-p)  \tilde f(p)  $ is rapidly decreasing  
\be
<F, f> = \lim_{N \to \infty}  <F, f_N>  =   \Vol(\bbT) ^{-1}\sum_{p } \tilde F(-p)  \tilde f(p)  
\ee
Finally    $F_N$ defined by 
\be
F_N =  \Vol(\bbT) ^{-1}  \sum_{ |p| \leq N} \tilde F( p)  e_p 
\ee
satisfies     $<F_N, f>  = <F, f_N>$ and  
is a smooth function which converges weakly to $F$.

  \newpage

\end{appendix}


\newpage

\begin{thebibliography}{BrDiHu95}




\bibitem{BDH95}  D. Brydges, J. Dimock, T. Hurd,  The short distance behavior of $(\phi^4)_3$,  Comm. Math. Phys. 172, (1995), 143-186. 


\bibitem{BDH96} 
 D.C.  Brydges, J. Dimock,  T. Hurd,  Estimates on renormalization group transformations, Canadian Journal of Mathematics 50 (1996), 756-793.


\bibitem{BDH98} 
 D.C.  Brydges, J. Dimock,  T. Hurd, A non-Gaussian fixed point for $\phi^4$ in 4-$\epsilon$ dimensions, Comm.  Math. Phys. 198 (1998), 111-156.

\bibitem{BrYa90}  D. C. Brydges,  H.T. Yau, Grad $\phi$ perturbations of massless Gaussian fields,  Commun. Math. Phys. 129,  (1990), 351-392. 

\bibitem{DiHu92} J. Dimock, T. Hurd,  A renormalization group analysis of correlations functions for the dipole gas,  J. Stat. Phys. 66, (1992), 1277- 1316.

\bibitem{DiYu24} J. Dimock, C. Yuan,  Structural stability of the RG flow in the Gross-Neveu model,  Annales Henri Poincar\'{e} 25,  (2024), 5113-5186.


\bibitem{FMRS86}  J. Feldman, J.Magnen, V. Rivasseau, R. Seneor,  A renormalizable field theory: the massive Gross-Neveu model in two dimensions,
Commun. Math. Phys 103, 67-103 (1986).


\bibitem{FKT00} J. Feldman, H. Kn\"{o}rrer, E. Trubowitz, 
Renormalization group and fermionic functional integrals, 
CRM Monograph Series, Volume 16, published by the AMS,  (2000). 


\bibitem{GrNe74} D. Gross, A Neveu, Dynamical symmetry breaking in asymptotically free theories,  Phys. Rev. D10, 3235-3253, (1974). 


\bibitem{GaKu84}  K. Gawedzki,  A. Kupiainen,  Lattice dipole gas and $(\nabla \phi)^4$ models at long distances: decay of correlations and scaling limit,
Commun. Math. Phys 92, 531-553,  (1984).


\bibitem{GaKu86}  K. Gawedzki,  A. Kupiainen,  Asymptotic freedom beyond perturbation theory,  in: K. Osterwalder, R. Stora, (eds.)  Critical Phenomena,
Random Systems, Gauge Theories,  Les Houches 1984, North Holland, (1986). 



\bibitem{GaKu85b}  K. Gawedzki,  A. Kupiainen,  Gross-Neveu model through convergent perturbation expansions, Commun. Math. Phys 102, 1-30 (1985).




\bibitem{Sim79}  B. Simon,     Trace ideals and their applications, Cambridge University Press, (1979).



\end{thebibliography}
\end{document}